\documentstyle[12pt]{article}
\def\A{{\cal A}}
\def\B{{\cal B}}

\def\D{{\cal D}}
\def\1{\parallel 1>}

\def\s{\sinh}

\begin{document}

\begin{titlepage}
\title{Exact Bethe Ansatz solution for $A_{n-1}$ chains with
non-$SU_{q}(n)$ invariant open boundary conditions.}

\author{H.J. de Vega \dag\thanks{Laboratoire Associ\'e au CNRS UA 280}\\
A. Gonz\'alez--Ruiz \ddag \thanks{Work supported by the Spanish M.E.C.
under grant AP90 02620085}\\
{\it \dag L.P.T.H.E.} \\
{\it Tour 16, 1er \'etage, Universit\'e Paris VI} \\
{\it 4 Place Jussieu, 75252 Paris cedex 05, FRANCE}\\
{\it \ddag Departamento de F\'{\i}sica Te\'orica} \\
{\it Universidad Complutense, 28040 Madrid, ESPA\~NA}}
\date{}
\maketitle

\begin{abstract}
The Nested Bethe Ansatz is generalized to open and independent boundary
conditions depending on two continuous and two discrete free
parameters.
This is used to find the exact eigenvectors and eigenvalues of
the  $A_{n-1}$ vertex models and $SU(n)$ spin chains
with such boundary conditions. The solution is found for all diagonal
families of solutions to the reflection equations in all possible
combinations. The Bethe ansatz equations are used to find de first order
finite size correction.
\end{abstract}

\vskip-16.0cm
\rightline{{\bf F.T/U.C.M-94/xx}}
\rightline{{\bf LPTHE--PAR 94/12}}
\rightline{{\bf March 1994}}
\vskip2cm

\end{titlepage}

As is by now well known integrability in the {\bf bulk} is a
consequence of the Yang-Baxter equation (YBE) in two-dimensional
lattice models and two-dimensional quantum field theory (QFT),
(see for example ref.\cite{rev}) .
Recently interest is focushed in generalizing boundary conditions (b.c.)
compatible with integrability. Periodic b. c. , twisted b. c. (by Cartan
algebra operators) and K-matrix b. c. in Sklyanin's framework provide
integrable models \cite{sk}. Several new families of b. c. compatible with
integrability has been recently found \cite{ks1,ks2,sz}.\\
The quantum group invariant chains can be obtained by
this method as special cases \cite{qba1,mn,ks1,ff}. In this construction the
reflection equations appear as the new ingredient to the bulk
Yang-Baxter equation
when dealing with open boundary conditions.
For the
moment only  models corresponding to rank one algebras has been solved with
independent open boundary conditions \cite{sk,lrv} and
recently the t-J model \cite{tj}. For
algebras of larger rank only  the quantum group invariant case of the
$A^{(2)}_2$, of the  $A_{n-1}$
and the t-J models has been solved \cite{lr,suq,fk}.\\
In this letter we solve the integrable models found in \cite{ks1}. They are
models with independent open boundary conditions associated to the algebras
$A_{n-1}$. We refer  to the article \cite{suq} for the notation and details of
some parts of the solution (we will work with a spectral parameter
shifted as
$\theta\rightarrow\theta+\gamma/2$ with respect to \cite{suq}).\\
The $A_{n-1}$ algebras have non symmetric $R$-matrices given in the hyperbolic
regime by:

\begin{eqnarray}
R^{ab}_{ab}(\theta)&=&\frac{\sinh\gamma}{\sinh(\theta+\gamma)}
e^{\theta\,{\rm sign}(a-b)}\;\;,\;\;a\neq b\nonumber\; \; ; \\
R^{ab}_{ba}(\theta)&=&\frac{\sinh\theta}{\sinh(\theta+\gamma)}\;\;, \;
\; a\neq b \; \; ;
\label{mran}\\
R^{aa}_{aa}(\theta)&=&1\nonumber\\
&&1\leq a,b\leq n\nonumber.
\end{eqnarray}

The diagonal solutions to the corresponding reflection equations were found
in \cite{ks1}, they are:

\begin{eqnarray}
&&{ }^{l_{-}}K^{-}_{a}(\theta)=
\frac{\s(\xi_{-}-\theta)}{\s\xi_{-}}\;e^{\theta}
\hspace{2cm} 1\leq a\leq l_{-}, \nonumber\\
&&{ }^{l_{-}}K^{-}_{a}(\theta)=\frac{\s(\xi_{-}+\theta)}{\s\xi_{-}}
\;e^{-\theta}
\hspace{2cm}l_{-}+1\leq a\leq n, \label{km}\\
&&{ }^{l_{+}}K^{+}_{a}(\theta)=\frac{\s(2\theta+\gamma)\;e^{(n-2a+1)\gamma}}
{\s(2\theta+n\gamma)}
\frac{\s(\xi_{+}+\theta)}{\s\xi_{+}}\;e^{-\theta}
\hspace{2cm} 1\leq a\leq l_{+}, \nonumber\\
&&{ }^{l_{+}}K^{+}_{a}(\theta)=\frac{\s(2\theta+\gamma)\;e^{(n-2a+1)\gamma}}
{\s(2\theta+n\gamma)}
\frac{\s(\xi_{+}-\theta-n\gamma)}{\s\xi_{+}}
\;e^{\theta+n\gamma}
\;\;l_{+}+1\leq a\leq n. \label{kp}
\end{eqnarray}
Notice that the quantum group invariant case is obtained in the limit
$\xi_{\pm}\rightarrow\infty$.\\
Using these solutions one can construct for an $A_{n-1}$ algebra, a set of
$(n-1)^{2}$ different transfer matrices depending on two arbitrary parameters
by taking  a ${ }^{l_{+}}K^{+}$ matrix and a ${ }^{l_{-}}K^{-}$ matrix. We
will call the transfer matrices constructed in this way
${ }^{l_{+}l_{-}}t(\theta,\tilde{\omega})$, where
$\tilde{\omega}=(\omega_{p_{0}},...,\omega_{1})$ are the first level
inhomogeneities of a chain of length $p_{0}$.\\
The way to diagonalize these transfer matrices is parallel to the quantum
group invariant case but we need som new ingredients. First we need to know
the action of the doubled monodromy matrix operators on the reference state.
The computation of this action is similar to the quantum group invariant case,
but now we use the expression (\ref{km}) (in the quantum group invariant case
$K^{-}(\theta)=1$). We have at level $(k)$ of the nested Bethe ansatz
construction:

\begin{eqnarray}
&&\A^{(k)}(\theta+(k-1)\gamma/2)\1={
}^{l_{-}}K^{(k)}_{1}(\theta+(k-1)\gamma/2) \1,\nonumber\\
&&U_{d1}^{(k)}(\theta+(k-1)\gamma/2)\1=0,\nonumber\\
&&\hat{\D}^{(k)}_{db}(\theta+(k-1)\gamma/2)\1=
\Delta^{(k)}_{d}(\theta+(k-1)\gamma/2)\delta_{db}\1,\nonumber\\
&&\hat{\B}^{(k)}_{d}(\theta)\1\;\neq\; 0,\nonumber\\
&&\Delta^{(k)}_{d}(\theta)=e^{2\theta+k\gamma}\prod_{j=1}^{p_{k-1}}
\frac{\s[\theta+\mu_{j}^{(k-1)}+(k-1)\gamma]\s(\theta-\mu_{j}^{(k-1)})}
{\s[\theta+\mu_{j}^{(k-1)}+k\gamma]\s(\theta-\mu_{j}^{(k-1)}+\gamma)}
\;\;{ }^{l_{-}}K^{(k+1)}_{d}(\theta+k\gamma/2),\label{det} \nonumber\\
&&{ }^{\;l_{-}}K^{(k+1)}_{d}(\theta+k\gamma/2)=e^{-k\gamma/2}
\frac{\s[(\xi_{-}+k\gamma/2)-\theta-k\gamma/2]}{\s\xi_{-}}
\;e^{\theta+k\gamma/2}
\hspace{1cm} k+1\leq d\leq l_{-}, \nonumber\\
&&{ }^{l_{-}}K^{(k+1)}_{a}(\theta+k\gamma/2)=e^{-k\gamma/2}
\frac{\s[(\xi_{-}+k\gamma/2)+\theta+k\gamma/2]}
{\s\xi_{-}}\;e^{-\theta-k\gamma/2}
\hspace{1cm}l_{-}+1\leq d\leq n, \nonumber\\
\label{krm}
\end{eqnarray}
where there are no summations over repeated indices and $\1$ is the reference
state of the corresponding level. It is easily seen that the matrix
${ }^{\;l_{-}}K^{(k+1)}(\theta+k\gamma/2)$ obeys the reflection equation for a
problem of weights $R^{(k+1)}(\theta+k\gamma/2)$ with indices from $k+1$ to
$n$.  Also the ${ }^{l_{+}}K^{+}(\theta)$ matrix obeys:

\begin{eqnarray}
&&{ }^{l_{+}}K^{+}_{d}(\theta)=\frac{\s(2\theta+\gamma)\;e^{-(k-1)\gamma}}
{\s(2\theta+k\gamma)}{ }^{l_{+}}K^{(k)}_{d}(\theta+(k-1)\gamma/2)
\;\;,1\leq k\leq n,\;k\leq d\leq l_{+}\nonumber\\
&&{ }^{l_{+}}K^{(k)}_{d}(\theta+(k-1)\gamma/2)=e^{(k-1)\gamma/2}
\frac{\s(2\theta+k\gamma)}{\s(2\theta+n\gamma)}\;
e^{[(n-k+1)-2(d-k+1)+1]\gamma}
\nonumber\\
&&\frac{\s[(\xi_{+}-(k-1)\gamma/2)+\theta+(k-1)\gamma/2]}{\s\xi_{+}}
\;e^{-\theta-(k-1)\gamma/2},\;\;k\leq d\leq l_{+}, \nonumber\\
&&{ }^{l_{+}}K^{(k)}_{d}(\theta+(k-1)\gamma/2)=e^{(k-1)\gamma/2}
\frac{\s(2\theta+k\gamma)}{\s(2\theta+n\gamma)}\;
e^{[(n-k+1)-2(d-k+1)+1]\gamma}
\nonumber\\
&&\frac{\s[(\xi_{+}-(k-1)\gamma/2)-\theta-(n-k+1)\gamma-(k-1)\gamma/2]}
{\s\xi_{+}}\;e^{\theta+(n-k+1)\gamma+(k-1)\gamma/2},\;l_{+}+1\leq
d\leq n. \nonumber \\
\label{krp}
\end{eqnarray}
The matrix ${ }^{l_{+}}K^{(k)}(\theta+(k-1)\gamma/2)$ obeys the reflection
equation for a problem of weights $R^{(k)}(\theta+(k-1)\gamma/2)$ with indices
running from $k$ to $n$. In the previous equation some unimportant
factors have been
kept to make this property clearer.\\
Using eq.(\ref{kp}), the transfer matrix can be written as:

\begin{eqnarray}
{ }^{l_{+}l_{-}}t(\theta,\tilde{\omega})&=&
\sum_{d=1}^{n}K^{+}_{d}(\theta)
U_{d}(\theta,\tilde{\omega})\nonumber\\
&=&\frac{\s[\xi_{+}+\theta-(n-l_{+})\gamma]}{\s\xi_{+}}\;
e^{-\theta+(n-l_{+})\gamma}
\A(\theta)\nonumber\\
&+&\frac{\s 2\theta}{\s(2\theta+2\gamma)}e^{-2\theta-\gamma}\;
\sum^{n}_{d=2}{ }^{l_{+}}K^{(2)}_{a}(\theta+\gamma/2)\;\hat{\D}_{dd}(\theta).
\label{tr}
\end{eqnarray}
%The operators $\hat{\D}$ of the previous equation are given by:
%\begin{eqnarray}
%\hat{\D}_{bd}(\theta)&=&\frac{1}{\s 2\theta}[e^{2\theta}\s(2\theta+\gamma)
%\D_{bd}(\theta)-\s\gamma\delta_{bd}\A(\theta)]
%\end{eqnarray}

Now we want to evaluate the action of the transfer matrix on the first level
Bethe Ansatz wave function, given by:

\begin{equation}
\Psi=\sum_{2\leq i_{j}\leq
n}X^{i_{1}\ldots i_{p_{1}}}\hat{\B}_{i_{1}}(\mu^{(1)}_{1})\ldots
\hat{\B}_{i_{p_{1}}}(\mu^{(1)}_{p_{1}})\1\nonumber\\
=\hat{\B}(\mu^{(1)}_{1})\otimes\ldots\otimes\hat{\B}(\mu^{(1)}_{p_{1}})X\1.
\label{bare}
\end{equation}
This is done using the commutation relations between the doubled monodromy
matrix operators, which are the same as in the quantum group invariant case.
The result is:

\begin{eqnarray}
&&{ }^{l_{+}l_{-}}t(\theta,\tilde{\omega})\Psi=\nonumber\\
&&\frac{\s[\xi_{+}+\theta-(n-l_{+})\gamma]\s(\xi_{-}-\theta)}
{\s\xi_{+}\s\xi_{-}}\;
e^{(n-l_{+})\gamma}
\prod^{p_{1}}_{j=1}
\frac{\s(\theta+\mu^{(1)}_{j})\s(\theta-\mu^{(1)}_{j}-\gamma)}
{\s(\theta+\mu^{(1)}_{j}+\gamma)\s(\theta-\mu^{(1)}_{j})}\Psi\nonumber\\
&&+\frac{\s(2\theta)}{\s(2\theta+2\gamma)}\prod_{i=1}^{p_{0}}
\frac{\s(\theta+\omega_{i})
\s(\theta-\omega_{i})}
{\s(\theta+\omega_{i}+\gamma)\s(\theta-\omega_{i}+\gamma)}
\label{want}\\
&&\prod^{p_{1}}_{j=1}
\frac{\s(\theta+\mu^{(1)}_{j}+2\gamma)\s(\theta-\mu^{(1)}_{j}+\gamma)}
{\s(\theta+\mu^{(1)}_{j}+\gamma)\s(\theta-\mu^{(1)}_{j})}\nonumber\\
&&\hat{\B}_{j_{1}}(\mu^{(1)}_{1})\ldots\hat{\B}_{j_{p_{1}}}(\mu^{(1)}_{p_{1}})
\1\;{ }^{l_{+}l_{-}}t^{(2)}(\theta;\tilde{\mu^{(1)}})^{j_{1}\ldots
j_{p_{1}}}_{i_{1}\ldots i_{p_{1}}} X^{i_{1}\ldots i_{p_{1}}}\nonumber\\
&&+~~u.t,\nonumber
\end{eqnarray}
where $ u.t $ are the unwanted terms. We are then led to a reduced problem of
transfer matrix ${ }^{l_{+}l_{-}}t^{(2)}(\theta;\tilde{\mu^{(1)}})$
with weights $R^{(2)}(\theta+\gamma/2)$, reflection matrices given by ${
}^{l_{+}}K^{(2)}(\theta+\gamma/2)$, ${
}^{\;l_{-}}K^{(2)}(\theta+\gamma/2)$ and inhomogeneities given by the first
level Bethe Ansatz. The reduced eigenvalue problem is:

\begin{eqnarray}
{ }^{l_{+}l_{-}}t^{(2)}(\theta;\tilde{\mu^{(1)}})X=
{ }^{l_{+}l_{-}}\Lambda^{(2)}(\theta;\tilde{\mu^{(1)}})X.
\end{eqnarray}

Using that $tr_{2}\;R^{(2)}_{12}(\theta+\gamma/2)
\;{ }^{l_{+}}K^{(2)}_{2}(\theta+\gamma/2)$
is a diagonal matrix is possible to
see that the unwanted terms cancel if:

\begin{eqnarray}
&&{ }^{l_{+}l_{-}}\Lambda^{(2)}(\mu^{(1)}_{k};\tilde{\mu^{(1)}})=
\frac{\s[\xi_{+}+\mu^{(1)}_{k}-(n-l_{+})\gamma]\s(\xi_{-}-\mu^{(1)}_{k})}
{\s\xi_{+}\s\xi_{-}}\;
e^{(n-l_{+})\gamma}\nonumber \\
&&\prod^{p_{1}}_{i\neq k}
\frac{\s(\mu^{(1)}_{k}+\mu^{(1)}_{i})\s(\mu^{(1)}_{k}-\mu^{(1)}_{i}-\gamma)}
{\s(\mu^{(1)}_{k}+\mu^{(1)}_{i}+2\gamma)\s(\mu^{(1)}_{k}-\mu^{(1)}_{i}+\gamma)}
\nonumber \\
&&\prod_{i=1}^{p_{0}}
\frac{\s(\mu^{(1)}_{k}+\omega_{i}+\gamma)
\s(\mu^{(1)}_{k}-\omega_{i}+\gamma)}
{\s(\mu^{(1)}_{k}+\omega_{i})\s(\mu^{(1)}_{k}-\omega_{i})},
\;\;1\leq k\leq p_{1}.
\end{eqnarray}
The analytic properties of the eigenvalues can be also used as a short way to
obtain these equations. One can follow this strategy level by level using
equations (\ref{krm},\ref{krp}). When the Bethe ansatz level is
$min(l_{+},l_{-})$ we can extract a common factor from the reduced problem
of order $min(l_{+},l_{-})+1$. The same occur when the level is
$max(l_{+},l_{-})$. In this case the reduced problem has the $K^{\pm}$
matrices of the quantum group invariant case. These factors are:

\begin{eqnarray}
&&\frac{\s(\xi_{-}+\theta+l_{-}\gamma)}{\s(\xi_{-})}\;e^{-\theta-l_{-}\gamma}
:=f_{l{-}},\;\;k=l_{-},
\nonumber\\
&&\frac{\s(\xi_{+}-\theta-n\gamma)}{\s(\xi_{+})}\;e^{\theta+n\gamma}
:=f_{l_{+}},\;\;k=l_{+},
\nonumber\\
&&\frac{\s(\xi_{+}-\theta-n\gamma)\s(\xi_{-}+\theta+l\gamma)}
{\s(\xi_{+})\s(\xi_{-})}\;e^{(n-l)\gamma}:=f_{l},\;\;k=l_{+}=l_{-}=l,
\end{eqnarray}
where $k$ is the Bethe ansatz level. Using the previous results the
recurrence relation for the eigenvalues can be obtained. The general
expression is:

\begin{eqnarray}
&&{ }^{l_{+}l_{-}}\Lambda^{(k)}(\theta,\tilde{\mu}^{(k-1)})=\nonumber\\
&&a^{(k)}(\theta)
\prod^{p_{k}}_{j=1}
\frac{\s[\theta+\mu^{(k)}_{j}+(k-1)\gamma]\s(\theta-\mu^{(k)}_{j}-\gamma)}
{\s(\theta+\mu^{(k)}_{j}+k\gamma)\s(\theta-\mu^{(k)}_{j})}\nonumber\\
&&+b^{(k)}(\theta)
\frac{\s[2\theta+(k-1)\gamma]}{\s[2\theta+(k+1)\gamma]}\prod_{i=1}^{p_{k-1}}
\frac{\s[\theta+\mu^{(k-1)}+(k-1)\gamma)
\s(\theta-\mu^{(k-1)})}
{\s(\theta+\mu^{(k-1)}+k\gamma)\s(\theta-\mu^{(k-1)}+\gamma)}
\label{rec}\nonumber\\
&&\prod^{p_{k}}_{j=1}
\frac{\s[\theta+\mu^{(k)}_{j}+(k+1)\gamma]\s(\theta-\mu^{(k)}_{j}+\gamma)}
{\s(\theta+\mu^{(k)}_{j}+k\gamma)\s(\theta-\mu^{(k)}_{j})}
\Lambda^{(k+1)}(\theta,\tilde{\mu}^{(k)}),\nonumber\\
&&1\leq k\leq n-1,\;\;\mu^{(0)}_{j}=\omega_{j},\;\;
\Lambda^{(n)}(\theta,\tilde{\mu}^{(n-1)})=1.
\end{eqnarray}

The expressions of the functions $a^{(k)}(\theta),b^{(k)}(\theta)$ are given
by:

\begin{eqnarray}
&&a^{(k)}(\theta)=a,\;\;\;b^{(k)}(\theta)=1,\;\;\;
1\leq k\leq min(l_{+},l_{-})-1,\nonumber\\
&&a^{(l_{<})}(\theta)=a,\;\;\;b^{(l_{<})}(\theta)=f_{l_{<}}
,\;\;\; k=min(l_{+},l_{-}):=l_{<},\nonumber\\
&&a^{(k)}(\theta)=g_{l_{>}},\;\;\;b^{(k)}(\theta)=1,
\;\;\;l_{<}+1\leq k\leq max(l_{+},l_{-})-1:=l_{>}-1,\nonumber\\
&& a^{(l_{>})}(\theta)=g_{l_{>}},\;\;\;b^{(l_{>})}(\theta)=f_{l_{>}},
\;\;\;k=l_{>},\nonumber\\
&&a^{(k)}(\theta)=b^{(k)}(\theta)=1,\;\;\;l_{>}+1\leq k\leq n-1,\label{eu}
\end{eqnarray}
where in the case $l_{+}=l_{-}=l$, $f_{l_{>}}=f_{l_{<}}=f_{l}$ and
$g_{l_{>}}=a$. In the previous expressions:

\begin{eqnarray}
&&a=\frac{\s[\xi_{+}+\theta-(n-l_{+})\gamma]\s(\xi_{-}-\theta)}
{\s\xi_{+}\s\xi_{-}}\;
e^{(n-l_{+})\gamma},\nonumber\\
&&g_{l_{+}}=\frac{\s[\xi_{+}+\theta-(n-l_{+})\gamma]}{\s\xi_{+}}\;
e^{-\theta+(n-l_{+})\gamma},\nonumber\\
&&g_{l_{-}}=\frac{\s(\xi_{-}-\theta)}
{\s\xi_{-}}\;
e^{\theta}.\label{ed}
\end{eqnarray}

We can now use the recurrence formula (\ref{rec}), and with the help of
equations (\ref{eu},\ref{ed}) find the expression for the eigenvalue of
${ }^{l_{+}l_{-}}t(\theta,\tilde{\omega})$. The result is:

\begin{eqnarray}
&&{ }^{l_{+}l_{-}}\Lambda(\theta,\tilde{\omega})=
\prod_{j=1}^{p_{0}}
\frac{\s(\theta+\omega_{j})\s(\theta-\omega_{j})}
{\s(\theta+\omega_{j}+\gamma)\s(\theta-\omega_{j}+\gamma)}\nonumber\\
&&\sum_{k=1}^{n}
g^{(k)}(\theta)\frac{\s(2\theta)\s(2\theta+\gamma)}
{\s[2\theta+(k-1)\gamma]\s[2\theta+k\gamma]}\nonumber\\
&&\prod_{j=1}^{p_{k-1}}
\frac{\s[\theta+\mu^{(k-1)}_{j}+k\gamma]\s(\theta-\mu^{(k-1)}_{j}+\gamma)}
{\s[\theta+\mu_{j}^{(k-1)}+(k-1)\gamma]\s(\theta-\mu_{j}^{(k-1)})}\nonumber\\
&&\prod_{j=1}^{p_{k}}
\frac{\s[\theta+\mu_{j}^{(k)}+(k-1)\gamma]\s(\theta-\mu_{j}^{(k)}-\gamma)}
{\s[\theta+\mu_{j}^{(k)}+k\gamma]\s(\theta-\mu_{j}^{(k)})}\label{autov},
\end{eqnarray}
where the product over $p_{n}$ is equal to one and
$\mu^{(o)}_{j}=\omega_{j}$. The value of $g^{(k)}(\theta)$ is given by:

\begin{eqnarray}
&&g^{(k)}(\theta)=a,\;\;\;1\leq k\leq l_{<},\nonumber\\
&&g^{(k)}(\theta)=f_{l_{<}}g_{l_{>}},\;\;\;l_{<}+1\leq k\leq l_{>},\nonumber\\
&&g^{(k)}(\theta)=f_{l_{+}}f_{l_{-}},\;\;\;l_{>}+1\leq k\leq n\label{g}.
\end{eqnarray}

The expression for the eigenvalue (\ref{autov},\ref{g}) gives the quantum
 group
invariant result in the limit  $\xi_{\pm}\rightarrow\infty$ \cite{suq}. For
the $SU(2)$
case this formulas reduce to those in reference \cite{sk}.
One can see that the
''mixed'' cases where $l_{+}\neq l_{-}$ give imaginary eigenvalues in the
trigonometric regime. For the cases when $l_{+}=l_{-}=l$ the eigenvalues are
essentially real, we can get rid of the exponential factor by a redefinition
of the  the reflection matrices $K^{+}\rightarrow e^{-i(n-l)\gamma}K^{+}$
\cite{tj}. This factor has been kept to make an easier connection with the
quantum group invariant case.\\
The Bethe ansatz equations that the roots $\mu^{(k)}_{i}$ have to obey are:

\begin{eqnarray}
h^{(k)}(\mu^{(k)}_{i})
\prod^{p_{k}}_{j\neq i}\frac{\s[\mu^{(k)}_{i}+\mu^{(k)}_{j}+(k-1)\gamma]
\s(\mu^{(k)}_{i}-\mu^{(k)}_{j}-\gamma)}
{\s[\mu^{(k)}_{i}+\mu^{(k)}_{j}+(k+1)\gamma]
\s(\mu^{(k)}_{i}-\mu^{(k)}_{j}+\gamma)}=\nonumber\\
\prod^{p_{k+1}}_{j=1}\frac{\s(\mu^{(k)}_{i}+\mu^{(k+1)}_{j}+k\gamma)
\s(\mu^{(k)}_{i}-\mu^{(k+1)}_{j}-\gamma)}
{\s[\mu^{(k)}_{i}+\mu^{(k+1)}_{j}+(k+1)\gamma]
\s(\mu^{(k)}_{i}-\mu^{(k+1)}_{j})}\nonumber\\
\prod^{p_{k-1}}_{j=1}\frac{\s[\mu^{(k)}_{i}+\mu^{(k-1)}_{j}+(k-1)\gamma]
\s(\mu^{(k)}_{i}-\mu^{(k-1)}_{j})}
{\s(\mu^{(k)}_{i}+\mu^{(k-1)}_{j}+k\gamma)
\s(\mu^{(k)}_{i}-\mu^{(k-1)}_{j}+\gamma)}\nonumber\\
1\leq k\leq n-1,\;\;1\leq i\leq p_{k}.
\label{lio}
\end{eqnarray}
In the previous expression the function $h^{(k)}(\theta)$ is given by:

\begin{eqnarray}
l_{+}\neq l_{-},\;\;\;h^{(l_{-})}(\mu^{(l_{-})}_{i})&=&
\frac{\s(\xi_{-}-\mu^{(l_{-})}_{i})}
{\s(\xi_{-}+\mu^{(l_{-})}_{i}+l_{-}\gamma)}
\;e^{2\mu^{(l_{-})}_{i}+l_{-}\gamma},
\nonumber\\
h^{(l_{+})}(\mu^{(l_{+})}_{i})&=&
\frac{\s[\xi_{+}+\mu^{(l_{+})}_{i}-(n-l_{+})\gamma]}
{\s(\xi_{+}-\mu^{(l_{+})}_{i}-n\gamma)}
\;e^{-2\mu^{(l_{+})}_{i}-l_{+}\gamma},
\nonumber\\
l_{+}=l_{-}=l,\;\;\;h^{(l)}(\mu^{(l)}_{i})&=&h^{(l_{-})}(\mu^{(l_{-})}_{i})
h^{(l_{+})}(\mu^{(l_{+})}_{i}),\nonumber\\
k\neq l_{+},l_{-},\;\;\;\;\;h^{(k)}(\theta)&=&1
\nonumber
\end{eqnarray}

{F}rom this last expressions we see that the eigenvalues of the transfer
matrix
in the trigonometric regime when $l_{+}=l_{-}=l$ are essentially real and
otherwise imaginary.\\

Let us now derive the solution of the present NBA equations in the
 thermodynamic limit for the ground state and in the gapless regime. For this
purpose, it is convenient to relate these equations with those of the
periodic case
 by means of the change of variables, (see \cite{qba1,suq}):

\begin{eqnarray}
\lambda^{(k)}_{s}&=&\nu^{(k)}_{s}\nonumber\\
\lambda^{k}_{2P_{k}-s+1}&=&-\nu^{(k)}_{s}
\label{cam}\\
1\leq s\leq p_{k}&;&1\leq k\leq n-1,\nonumber
\end{eqnarray}
where $\nu^{(k)}_{j}=-i\mu^{(k)}_{j}-ik\gamma/2$. To pass to the gapless
regime we make also the changes $\xi_{\pm}\rightarrow -i\xi_{\pm}$,
$\gamma\rightarrow -i\gamma$, $\nu^{(k)}_{j}\rightarrow i\nu^{(k)}_{j}$. Then,
following standard techniques \cite{rev} one introduces a density  of roots
at every NBA level (from now on $p_{0}=N$):

\begin{eqnarray}
\rho^{l}(\lambda^{(l)}_{j})=\lim_{N\rightarrow\infty}
\frac{1}{N(\lambda^{(l)}_{j+1}-\lambda^{(l)}_{j}))}\nonumber
\end{eqnarray}

As explained in refs.\cite{qba1,suq} a hole at $\lambda = 0$
shows up in the ground state. Then

\begin{eqnarray}
\sigma^{l}(\lambda) = \rho^{l}(\lambda) + {1 \over N} \delta(\lambda)
\nonumber
\end{eqnarray}
is a regular and continuous function. In the $N \to \infty$
limit, eqs.(\ref{lio}) become through standard techniques a system of
integral equations for the densities $\sigma^{l}(\lambda)$.

\begin{eqnarray}
&&\sigma^{k}(\lambda)-\sum_{m=1}^{n-1}\int_{-\infty}^{\infty}
d\mu K_{km}(\lambda-\mu)\sigma^{m}(\mu)\nonumber\\
&&=\frac{1}{2\pi N}\;\Phi^\prime(\lambda,\gamma/2)-\frac{1}{2\pi
N}\;\Phi^\prime(\lambda,(\pi-\gamma)/2)
\nonumber \\
&&+\frac{\delta_{k1}}{\pi}\;\Phi^\prime(\lambda,\gamma/2)-
\frac{1}{N}\sum_{m=1}^{n-1}K_{km}(\lambda)
+ \frac{1}{2\pi N}\;\chi'_k(\lambda)
\label{sei}\\
&&\chi_k(\lambda)= -i\log h^{(k)}(\lambda)
\end{eqnarray}
That is,
\begin{eqnarray}
&&\chi_{l_-}(\lambda)= 2i\lambda +
\Phi(\lambda, \xi_- + \frac{\gamma}{2}l_-) \,,\,
\chi_{l_+}(\lambda)= - 2i\lambda -
\Phi(\lambda, \xi_+ - \gamma (n -\frac{l_+}{2}))
\nonumber \\
&&\chi_k = 0 {\rm ~when~} k  \neq l_+,~l_- .\\
&&{\rm When~}~  l_+ = l_- = l,~~
\chi_{l}(\lambda)=\chi_{l_-}(\lambda)+\chi_{l_+}(\lambda)
\end{eqnarray}
Here

\begin{eqnarray}
\Phi(z,\gamma)=i\; \log\left[\frac{\s(i\gamma+z)}{\s(i\gamma-z)}\right]
\nonumber
\end{eqnarray}

When $ \chi_k(\lambda)=0 $ we recover the $SU_{q}(n)$ invariant case
solved in ref.\cite{suq}. The density of real roots in our case
results  equal to $\sigma^{l}(\lambda)$ in ref.\cite{suq} plus
the term
\begin{eqnarray}
&&\delta\sigma^{l}(\lambda)=
\int_{-\infty}^{\infty}\frac{dk}{2\pi}e^{ik\lambda}
\; \delta\hat{\sigma}^{l}(k)\\ \nonumber
&&\delta \hat{\sigma}^{l}(k) = \frac{1}{N} \left[ \sinh{k(\frac{\pi}{2}
- \eta_-)} \hat{R}_{l l_-}(k) - \sinh{k(\frac{\pi}{2}
- \eta_+)} \hat{R}_{l l_+}(k) \right]
\frac{1}{\sinh{k\frac{\pi}{2}}} \nonumber \\
&& +\frac{2 i}{N} \delta(k)\left[  \hat{R}_{l l_-}(0) -  \hat{R}_{l l_+}(0)
\right] \nonumber
\end{eqnarray}
One can see that in the quantum group invariant case this term goes to zero.
Here $\hat{R}_{ll^\prime}(k)$ stands for the resolvent kernel:
\begin{eqnarray}
\hat{R}_{ll^\prime}(2x)=\frac{\s(\pi x)\s[\gamma x(n-l_{>})]\s(\gamma
x l_{<})}{\s[x(\pi-\gamma)]\s(\gamma x n)\s(\gamma x)}
\label{resok}\nonumber
\end{eqnarray}
and
\begin{eqnarray}
\eta_- = \xi_- + \frac{\gamma}{2}l_- ~~,~~
\eta_+= \xi_+ - \gamma (n -\frac{l_+}{2}) \nonumber
\end{eqnarray}
(We assume $ \pi > \eta_{\pm} > \gamma/2 $).
We will see what is the change in the free energy for the case when
$l_{+}=l_{-}=l$, in the other cases one obtains imaginary
 contributions due
to the fact that the eigenvalues are imaginary, we also perform the
redefinition of the $K^{+}$ matrix mentioned above. The change on the free
energy compared to the $SU(n)_q$ invariant case is:

\begin{eqnarray}
&&\delta f(\theta) =  \frac{4}{N} \int_{0}^{\infty}\frac{dk}{k}~
\frac{\sinh k\theta\;\sinh[k\gamma(n-l)/2]\;\sinh[k(\eta_{+}-\eta_{-})/2]
\;\cosh[k(\pi-\eta_{+}-\eta_{-})/2]}
{\sinh{(\gamma n k / 2)} \sinh{(\pi k/2)}}
\nonumber \\
%&&\left[ \sinh{k(\frac{\pi}{2} - \eta_-)} \sinh{\frac{k \gamma}{2}(n-l_-)} -
%\sinh{k(\frac{\pi}{2} - \eta_+)} \sinh{\frac{k \gamma}{2}(n-l)}\right]
%\nonumber\\
&&-\frac{1}{N}\log\left\{\frac{\sin[\xi_{+}+\theta-(n-l)\gamma]}
{\sin \xi_{+}}\right\}-\frac{1}{N}\log\left\{\frac{\sin(\xi_{-}
-\theta)}
{\sin \xi_{-}}\right\}\nonumber
\end{eqnarray}

This term represents the surface energy induced by our general
boundary conditions as a function of $\xi_+$ and  $\xi_-$.
The derivative of $\delta f(\theta)$ at $\theta = 0$ times
$ -\frac{\sin \gamma}{2} $ provides the surface change on the ground
state of the associated magnetic chain.\\
In conclusion we have obtained the expression for the eigenvalues and nested
Bethe ansatz equations for the open $A_{n-1}$ models of reference \cite{ks1}.
For some regimes the ''mixed'' cases give complex  eigenvalues. For the cases
 where $l_{+}=l_{-}=l$ the eigenvalues are real for all the regimes even
though the hamiltonians are not hermitian. We have obtained the first order
finite size corrections in this case. It would be interesting to study the
thermodynamic properties of these chains. In the other hand these
hamiltonians, even in the mixed cases, could have some application in the
framework developed in \cite{ritt} for the reaction diffusion equations.

\vspace{2cm}
AGR would like to thank L. A. Ibort for comments.

\end{document}